\documentclass[preprints,review,accept,oneauthor,12pt]{Definitions/mdpi} 

\firstpage{1} 
\pubvolume{7}
\issuenum{3}
\articlenumber{75}
\pubyear{2019}
\copyrightyear{2019}
\history{Received: 24 July 2019; Accepted: 12 August 2019; Published: 23 August 2019}

\usepackage{subcaption}
\usepackage{geometry}
\usepackage{pdflscape}

\Title{The Complex Upper HR Diagram}

\Author{Roberta M. Humphreys $^{1,\dagger}$\orcidA{0000-0003-1720-9807}}

\AuthorNames{Firstname Lastname, Firstname Lastname and Firstname Lastname}

\address{ $^{1}$ \quad University of Minnesota ; roberta@umn.edu}  

\corres{Correspondence: roberta@umn.edu; Tel.: +1-612-624-6530}

\firstnote{Current address: Minnesota Institute for Astrophysics, University of Minnesota, 116 Church St. SE, Minneapolis, MN 55455} 

\abstract{Several decades of observations of the most massive and most luminous stars 
have revealed a complex upper HR Diagram, shaped by mass loss, and inhabited by a 
variety of evolved stars exhibiting the consequences of their mass loss histories. This Introductory chapter presents a  brief historical overview of the HR Diagram for massive 
stars, highlighting some of the primary discoveries and results from their observation in nearby galaxies. The chapters in this volume include reviews of our current understanding of different groups of evolved massive stars, all losing mass and in different stages of their 
evolution; the Luminous Blue Variables (LBVs), B[e] supergiants, the warm hypergiants, Wolf-Rayet stars, and the population of OB stars and supergiants in the Magellanic Clouds. }


\begin{document}


\section{Introduction}

The reviews and papers in this volume focus on the properties of the most luminous 
stars in nearby galaxies, those galaxies in which the brightest individual 
stars are resolved and can be observed. The most luminous stars are also the 
most massive and because of their intrinsic brightness and relatively short 
lifetimes they provide our first probes of the progress of stellar evolution in 
different environments.   

The study of massive stars in other galaxies offers many advantages. Foremost of
course is distance. Studies of stellar populations in nearby galaxies  have the 
advantage that all the stars are at approximately the same distance, 
a distance that is relatively well known, especially in comparison with the 
uncertain distances of individual stars in our own galaxy. In the Milky Way, our
 observations are also limited to a relatively small volume by interstellar 
 extinction which can be high and uncertain at increasing distances.  In external
  galaxies, extinction by dust is still a problem but the foreground extinction 
  is well determined from maps of the interstellar ``cirrus'' or dust along the line 
  of sight. Internal extinction within the galaxy  can be variable and 
  must still be corrected. 

In this introductory chapter, I present a brief historical overview with 
emphasis on some of the main developments in the study of massive stars in 
nearby galaxies.  Some of the first work on stars in other galaxies was 
driven by the identification of Cepheids for the extragalactic distance scale, see for example, the
comprehensive survey of NGC 2403 by \citet{TM}.  
Other types of variables were also recognized including the first discussion 
of a class of  luminous variables by \citet{HS}  in the Local Group 
spirals M31 and M33 that we now call Luminous Blue Variables (LBVs). 
This work was done on photographic plates and the magnitudes of individual 
stars were often measured by hand.

The study of individual stars and their placement on the HR Diagram requires
several types of data, accurate multicolor photometry for spectral energy 
distributions (SEDs) and the measurment of interstellar extinction, infrared observations of 
circumstellar dust and mass loss but most important, spectroscopy,  for 
classification, temperatures, luminosity indicators, emission lines, and evidence for mass loss from 
line profiles. And when observing stars in other galaxies, we also have to be 
concerned about foreground contamination; stars in the Milky Way  seen 
projected against the distant galaxy. Photometry is not sufficient to remove 
these Galactic stars, especially those of intermediate temperature and late spectral type; spectra are required. 

The primary galaxies for the chapters in this volume 
are the Local Group spirals M31 and M33, and the Large and Small Magellanic Clouds due to
 their relative proximity and the number of surveys of their luminous 
 stellar populations. Other Local Group galaxies are also included as well as 
 results for stars in a few nearby spiral galaxies. 

\section{Early Work -- The Magellanic Clouds}

The early objective prism surveys by the Harvard College Observatory and the resulting Henry Draper Catalog and Extension (HD and HDE) provided the first survey with spectral types of the brightest stars in the Magellanic Clouds \citep{HA,Cannon}.  The HDE however did not cover the Small Cloud. 
\citet{Feast} published the first detailed list of the brightest stars 
in the Magellanic Clouds with spectral classifications based on slit spectra of individual stars. 
 Their paper included  magnitudes, colors, positions, radial velocities and notes on the individual stars relative to emission lines and other features for 50 stars in the SMC and 105 in the LMC. The HDE catalog was the primary source for the LMC stars, plus stars selected as blue based on their colors in the 30 Dor region. The SMC list relied on ``B'' type stars from the HD plus emission line stars from the survey by \citet{Henize}. Their work did not include the red or M-type supergiants. The reddest stars in their study were  the very luminous F and G-type supergiants. Their HR Diagram, reproduced here in Figure 1, while not complete, shows the visually brightest stars with M$_{v}$ approaching -10 mag. 

\begin{figure}
\centering
\includegraphics[width=11.5 cm]{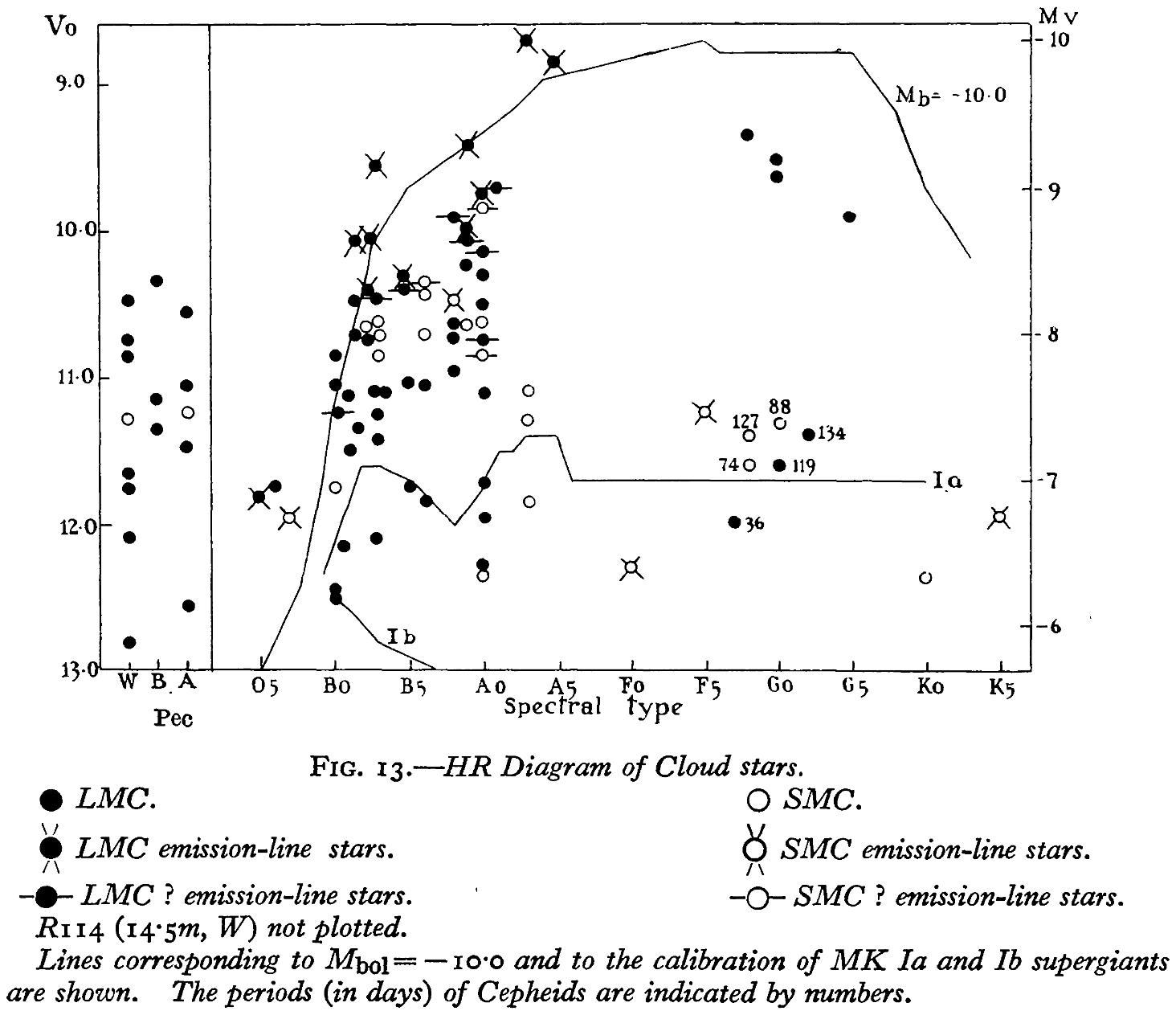}
\caption{The HR Diagram for the brightest stars in the LMC and SMC from \citet{Feast}. LMC stars are closed smbols and those in the SMC are shown as open symbols.}
\end{figure} 

Additional work on the fainter stellar populations of hot or OB-type stars and red M-type stars was provided by the lower resolution objective prism surveys throughout the 1960's and 1970's. These included surveys for OB type stars in the LMC and SMC by Sanduleak \cite{Sanduleak68,Sanduleak70} and by \citet{Fehrenbach}  and 
for the M type stars by Westerlund \cite{West60,West61}, \citet{Blanco80} and \citet{Blanco83}. These necessary photographic surveys provided extensive finding lists for further spectral classification leading to 
the first comprehensive 
HR diagrams for these two galaxies and the comparison of their massive  and luminous stellar populations with the Milky Way.

\section{The HR Diagram} 

Fundamental data for the most luminous stars have application for numerous astrophysical questions 
specifically with respect to stellar evolution, the final stages of the most massive stars as the 
progenitors of supernovae, and the dependence of their basic parameters on the host galaxy. In the 1970's there was also considerable interest in the luminosity calibration of the brightest stars and their 
potential  as extragalactic distance indicators \citep{ST74}. 

The massive star population in the Milky Way, although restricted to a relatively small volume, within $\approx$ 3 kpc of the Sun,   is  
critical as a reference population and as representative of our local region of the Galaxy, 
 despite  uncertain distances and 
possible incompleteness. \citet{RMH78} published an HR Diagram for the Galactic supergiants and O stars with 
spectral types and photometry in 
 stellar associations and clusters with known distances to derive  their luminosities. For a more 
complete population in the upper HR diagram, the less luminous early B-type main sequence and giant stars were also included. Compared to previous work,  numerous surveys, especially of the Southern sky, had added greatly to the number of confirmed supergiants and O stars \citep{RMH73,RMH75,Walborn72} and red supergiants 
\citep{RMH72,RMH74}. For comparison with 
evolutionary tracks, the derived absolute visual magnitudes and spectral types were transformed to
absolute bolometric luminosities and effective temperatures based on the available calibrations.    

\citet{RMH79} subsequently published HR Diagrams for the LMC.  The basic data for the 
confirmed 
supergiants and early-type stars came from the extensive catalogs of spectral types and photometry
by \citet{Ardeberg} and \citet{Brunet}, from \citet{Feast},  and 
from \citet{Walborn77} for the early O-type stars. The data for the M-type supergiants came from the spectroscopic survey in the same paper. 

An empirical comparison of these HR Diagrams, for the massive stars in our region of the Milky Way and 
for the LMC, revealed comparable populations of massive stars based on the distribution of their spectral types and luminosities across the HR Diagrams.  The most important result was the recognition by \citet{HD79}  
of an empirical upper luminosity boundary or upper limit in the luminosity vs temperture diagrams.  

\subsection{The Humphreys-Davidson Limit} 

The original HR diagrams from the Humphreys and Davidson 1979 paper are reproduced here in Figures 2 and 3.  
with the upper boundary shown as a solid line. The original eyeball fit was first drawn to approximate the upper boundary of the supergiant luminosities in the Milky Way. The same line was  then transferred 
to the LMC diagram which  also matched the observed upper envelope to the LMC luminosities.  The
upper boundary for both galaxies is an envelope of declining luminosity and decreasing temperature for the hottest stars and a relatively tight upper limit to the luminosities of the cooler stars (less than $\approx$ 10,000K) near M$_{Bol}$ $\approx$ -9.5 Mag. 

\begin{figure}
\centering
\includegraphics[width=\textwidth]{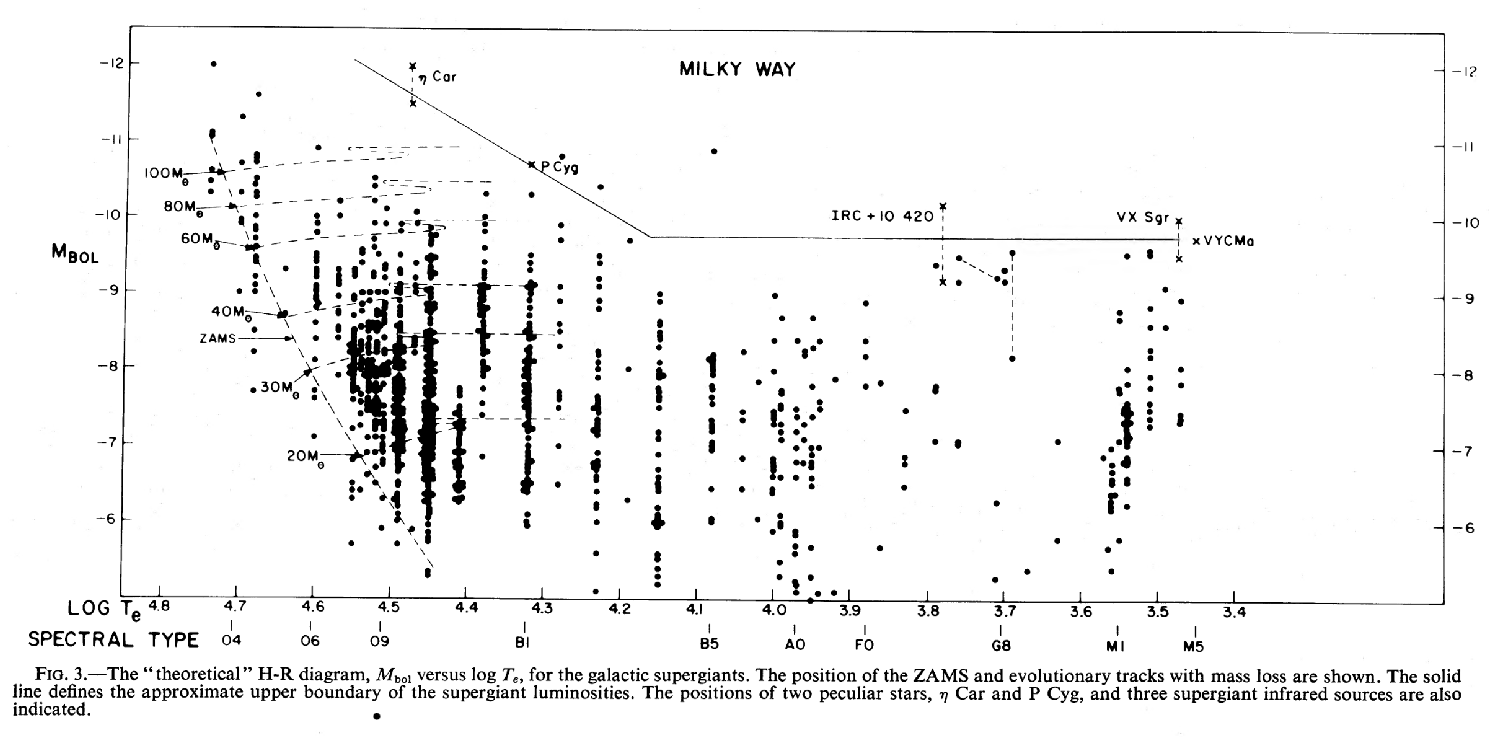}
\caption{The HR Diagram, M$_{Bol}$ vs. log T,  for the luminous  stars in the Milky Way from \citet{HD79}. }
\end{figure}

\begin{figure}
\centering
\includegraphics[width=\textwidth]{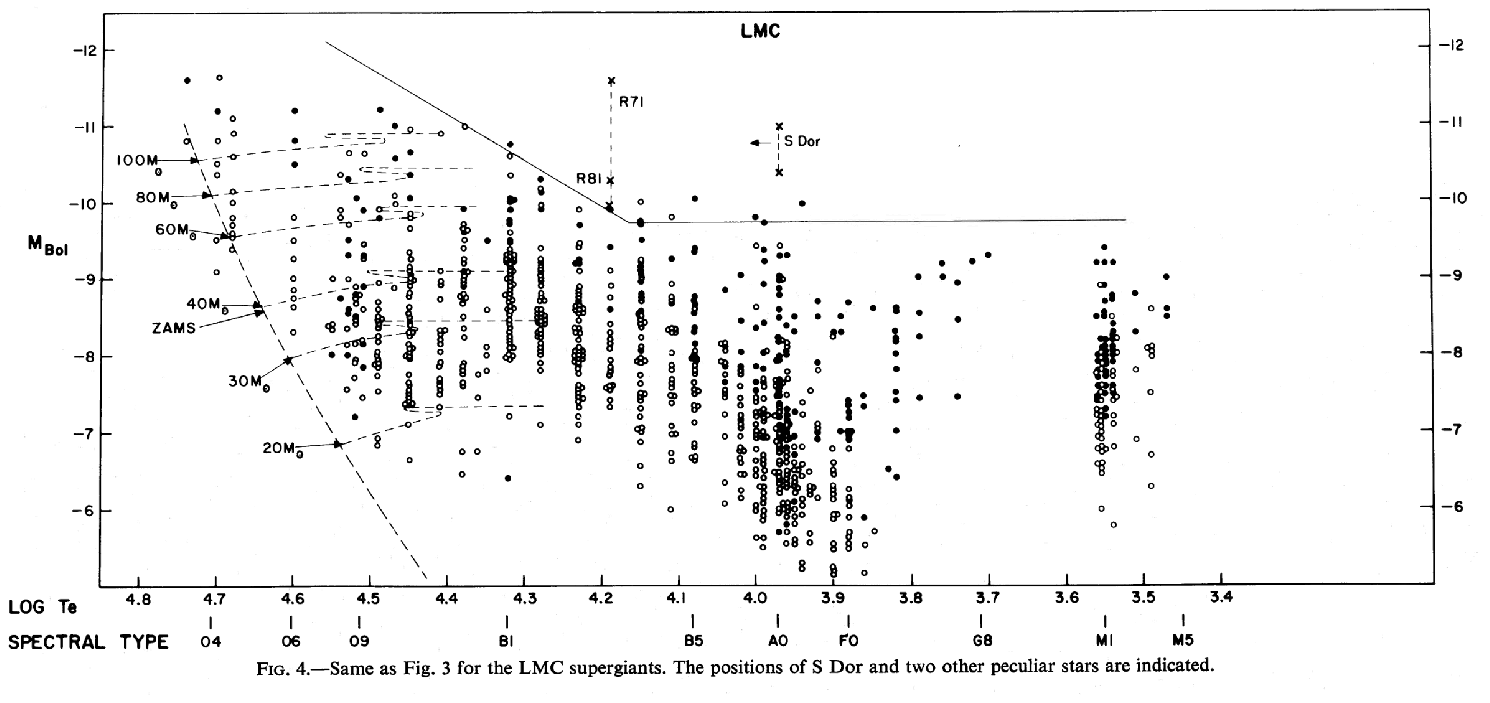}
\caption{The HR Diagram, M$_{Bol}$ vs. log T,  for the luminous  stars in the LMC from \citet{HD79}. }
\end{figure}

This empirical boundary is often referred to 
 in the literature as the ``Humphreys-Davidson'' Limit. It was not predicted by theory or the stellar 
structure models and evolutionary tracks at that time. The lack of evolved stars, post main sequence stars, above a certain luminosity implies an upper limit to the masses of stars that can evolve to become 
red supergiants thus altering the previously expected evolution of the most massive stars across the HR Diagram. In the 
original study, this limit corresponded to an initial mass near 60 M$_{\odot}$. Improved models with 
mass loss and rotation suggest a mass more like 40 - 50 M$_{\odot}$ today. 

Spectroscopy of luminous star candidates in other Local Group galaxies confirmed the upper 
luminosity boundary in galaxies of different types, but in most cases surveys and population studies were minimal; M31 \citep{RMH79b}, IC 1613 and NGC 6822 \citep{RMH80a}. \citet{HS80} completed a major survey for the brightest blue and red stars in M33, that provided the basis for spectroscopy and identification of the most luminous supergiants in that nearby spiral \citep{RMH80b}, and in subsequent studies.   
Similar HR Diagrams for the massive stars in the SMC were published a few years later by \citet{RMH83} based on a combination of previous  work and new observations.  

At the time the Humphreys-Davidson paper was published, it was  apparent that significant mass loss 
occurred in  both the blue and red supergiants and there was increasing interest in the role that mass 
loss may have on their evolution. The lack of evolved cooler counterparts to the most massive evolved hot stars in both galaxies suggested that their post main sequence evolution was of 
special interest.  A few high luminosity stars were  known for their instabilities, variability and evidence for high 
mass loss such as  eta Car and P Cyg in the Milky Way and S Dor in the LMC. A number of luminous blue variables, spectroscopically similar to eta Car and P Cyg, were now recognized in other galaxies \citep{RMH75_HS,RMH78_HS}. As a group, they were known as S Doradus  Variables or as Hubble-Sandage Variables in M31 and M33. Several of these stars 
were included on the HR Diagrams. They were increasing evidence that the phenomena of high mass loss and 
instabilities, as observed in eta Car and P Cyg, were more common. Humphreys and Davidson thus 
suggested that the most massive hot stars could not evolve to cooler temperatures because of their 
instabilities resulting in  high mass loss. The temperature dependence of the luminosity limit for the 
hottest stars was evidence that the instability was mass dependent. This mass loss could be unsteady and much greater at times resulting in high mass loss events. The relatively tight upper luminosity boundary for the cooler stars represented the upper limit to the initial masses of stars that could evolve  
across the HR Diagram in a stable way to become red supergiants (RSGs). 

Some of the first evolutionary tracks with mass loss were also being published at that time \citep{deLoore,Chiosi,Maeder80,Maeder81}. It was shown that high mass loss (higher than observed) would cause the tracks to reverse and
the stars evolve back to warmer temperatures. 

\section{Surveys and More Surveys}

We emphasize the importance of spectroscopy for the identification and analysis of the  most luminous stars, but  surveys are essential for identifying
candidates. Beginning in the 1980's astronomers  began to add multi-wavelength surveys, primarily  in the near-infrared, to complement the traditional optical photometry. \citet{Elias} obtained near IR photometry for known and candidate red supergiants in the LMC and SMC for a comparison with the Galactic population.  The all-sky near-infrared 
2MASS survey from 1.2  to 2.2$\mu$m \citep{Skrutskie} reached the brightest stars in 
M31 and M33 as well as the Clouds. The infrared observations allowed astronomers to look for free-free emission from the stellar winds and the presence 
of circumstellar dust, another  indicator of mass loss, and to correct the
luminosites for possible additional  extinction due to circumstellar dust. 

Space-based 
telescopes such as UIT and later GALEX added FUV and NUV  
imaging and photometry  for more complete SEDS at the shorter wavelengths and more accurate 
estimates  of the total luminosities for the hottest supergiants. 
The mid-infrared surveys with Spitzer/IRAC of the Magellanic Clouds \citep{Bonanos09,Bonanos10} and M31\citep{Mould} and M33\citep{McQuinn} added fluxes  
from 3 to 8$\mu$m and even longer wavelengths  to search for colder dust thus allowing us to investigate their mass loss histories.  

With the advent of the wide-field 
CCD mosaic cameras, groundbased, multi-wavelength optical surveys, such as the Local Group (LGGS) by \citet{Massey}, added to the fundamental data. These groundbased surveys however are seeing-limited and lack spatial resolution so 
the images are often multiple. Here again specroscopy or higher resolution 
imaging  with Adaptive Optics on large telescopes can identify and even separate the stars.  These 
 groundbased surveys are enhanced by observations with the Hubble Space 
 Telescope such as the PHAT surveys in M31 and M33 \citep{Dalcanton}. Numerous 
 imaging programs of other galaxies intended for other purposes with HST have 
 provided lists of resolved stars for further observation in, for example, 
 M101 \citep{Grammer} and NGC 2403 and M81 \citep{RMH2019,Kud}. 

\section{Stellar Population Comparisons} 

One of the outstanding stellar evolution questions is how  
stellar populations may depend on their environment; the properties of 
the host galaxy, its mass and luminosity, the fraction of  interstellar 
gas and dust, and especially on the chemical composition or metallicity.  
As the most luminous and visually brightest stars, massive stars provide the 
first indicator or measurement of how star formation, evolution, and 
the terminal state, depend on these factors. Mass loss is known to alter 
stellar evolution, and in the standard picture of line-driven winds from hot stars, mass loss is also  expected to be metallicity dependent and significantly less in lower 
metallicity systems.  

The first evidence for significant differences in the properties of the 
massive star populations was the well-established absolute magnitude 
dependence  of the visually ``brightest blue  star'' on the Galaxy type or luminosity in the surveys for 
extragalactic distance indicators \citep{ST74}. The smaller, less massive galaxies had fewer of the most massive stars thus their evolved counterparts  were 
statistically less likely to become the most  luminous observed blue stars. 
 These results suggested that  the star formation rate 
for these most massive stars was less in the smaller galaxies. In contrast, the 
luminosities of the brightest red stars showed little or no dependence on the 
host galaxy. The stars of somewhat lower initial mass thus existed in 
sufficient 
numbers in the smaller galaxies to produce evolved descendents in the  red supergiant region, and the luminosities of their  brightest members reflected the 
upper luminosity boundary. 

The metallicity dependence of the overall characteristics of the massive star 
poplulations in different galaxies has been less apparent. Comparison of the HR Diagrams of the massive stars in our region of the Milky Way with the LMC revealed very similar populations \citep{HD79}. The LMC oxygen abundance is lower than Solar 
but by no more than a factor of two. Consequently, a comparison with the 
outer regions 
 of the Milky Way, also with reduced  metallicity, not surprisedly,  showed little variation. Among our Local Group galaxies with comprehensive stellar surveys for luminous stars, 
 the SMC has the lowest metallicity, about 1/10th Solar, and a reduced oxygen 
 abundance by about a factor of five. 

Early tell-tale evidence 
for an observable metallicity affect, was the distribution of the spectral types of the M 
supergiants \citep{RMH_RSG}. A preliminary survey of red supergiants in the LMC and SMC compared to known Galactic RSGs, revealed a dramatic shift in their 
 spectral types in the SMC to much earlier spectral types, compared to the LMC 
 and Milky Way stars. Except for one M2 -type star in the SMC, all of the others
 were  type M0 or earlier. This apparent shift, attributed to weaker TiO bands, was due to the lower SMC metallicity resulting in lower opacities in the 
 atmospheres. The spectra thus arise in warmer layers. This result was confirmed in later surveys that extended to fainter magnitudes \citep{Elias,MasseyOL}.  

Stellar wind theory predicts a measurable dependence of the mass loss rate
on metallicity \citep{Vink}, decreasing with declining heavy element abundances, but the measured rates \citep{Mokiem} are somewhat higher than expected. Clumping in the stellar winds is another complication which when included in the 
mass loss models reduces the mass loss rates \citep{Crowther,Hillier,Fullerton}.  Measurement of the stellar wind properties and mass loss rates in the 
luminous, hot OB-type stars in nearby galaxies, especially in the Magellanic Clouds, has
progressed with the advent of very large telescopes equipped with high resolution spectrographs.  The chapter in this volume by Paul Crowther on the FLAMES survey the luminous, hot stars in the Clouds will discuss their winds and mass loss rates.

\subsection{The Most Luminous Stars of Different Types}

Since the early work of the 1970's and 80's to identify the most luminous and 
brightest stars in nearby galaxies, numerous surveys and studies of the massive stars, primarily in Local Group galaxies, have greatly expanded the 
completeness 
of the population samples. These include, in the LMC and SMC: surveys for the yellow and red supergiants \citep{Neugent2010,Neugent2012}, and in M31 and M33: studies 
of the luminous star population \citep{RMH2013,RMH2014,Massey2016b,RMH2017a,RMH2017b} and surveys for the yellow and red 
supergiants \citep{Drout2009,Massey2009,Drout2012,Gordon,Massey2016a}. Surveys and 
follow-up  spectroscopy of the luminous blue and red stars in the Local Group
irregulars NGC 6822 and IC 1613 are less complete \citep{RMH80a,Massey95,Massey98,Bresolin07}, but they provide an additional sample of the massive stars in two smaller galaxies and also with reduced metallicity.  

Table 1 presents  a summary of the most luminous stars of different spectral 
types or temperature ranges in six Local Group Galaxies with their morphological types and integrated visual luminosities.  Initial samples of the 
massive stars in the well-studied nearby spirals, M101, M81 and NGC 2403, 
outside our Local Group, have also been observed \citep{Zickgraf,Grammer,RMH86,RMH2019,Kud} and 
are included here. The bolometric luminosities of the stars are listed for the three highest in each spectral type group for each galaxy.  The 
adopted distance moduli (Table 2)  were used to determine the luminosities. The O-type stars are not included because many are eventually recognized as binary or are in multiple systems. Likewise, those  stars  in extremely crowded fields, including many OB-type stars, are not listed since they may be blended. Consequently, this Table does not include those stars that will be the intrinsically most luminous, most massive members of their home galaxies.  
A few stars of special interest are identified by name in the Table. Some of them will be discussed in later chapters in this volume. Note that the Galactic stars are not included. The upper HR Diagram for the Milky Way needs to re-examined when the Gaia survey is complete 
with improved distances and very likely with a larger volume sample.


\begin{table}[H]  
\caption{The Most Luminous Stars (M$_{Bol}$) of Different Spectral Types}
\centering
\tablesize{\footnotesize}
\begin{tabular}{llllllllll}
\toprule
\textbf{Galaxy} &  \textbf{M31} & \textbf{M33} & \textbf{LMC} & \textbf{SMC} & \textbf{NGC 6822} & \textbf{IC 1613} & \textbf{NGC 2403} &  \textbf{M81} &  \textbf{M101}  \\  
\textbf{Type}  &  \textbf{Sb I-II} & \textbf{Sc II} & \textbf{Im} & \textbf{Im} &   \textbf{Im}   & \textbf{Ir}  &  \textbf{Sc III} &  \textbf{Sb I-II} &  \textbf{Sc I}   \\ 
\textbf{M$_{v}$} &  \textbf{-21.5} & \textbf{-18.9} & \textbf{-18.5} & \textbf{-16.8} & \textbf{-15.2}   & \textbf{-14.8} &  \textbf{-19.3} &  \textbf{-20.9} &  \textbf{-21.0}   \\
\midrule
Spectral Type  &  M$_{Bol}$ & M$_{Bol}$ & M$_{Bol}$ & M$_{Bol}$ &  M$_{Bol}$ & M$_{Bol}$  & M$_{Bol}$ &  M$_{Bol}$ & M$_{Bol}$ \\
O9.5 -- B5  & -10.6(2)  & -10.1 & -10.9 & -10.3(2) & -10.3    &  -9.5  & -10.0(2)   & -10.6  &  ... \\ 
            & -10.2     & -10.0 & -10.7 & -10.2     & -9.9    & -8.8  &  -9.5  & -10.0  & ... \\
	    & -10.0(2)  & -9.8  &  -10.0 & -9.8(2)   &  -9.5    &  -8.7  &  -9.4  & -9.7  & ...  \\
	    &           &       &        &       &        &    &    &   &  \\
B8 -- A8    &  -9.3     & -10.1(B324) &  -9.6(HD33579)  &  -9.6(HD7583)  & -8.1  & -8.4  & -9.8   & -9.3  & -10.4(2)\\
            &  -9.1    & -9.5(2)     & -9.0       &  -9.2   &  -7.7      &  -6.9 & -9.5   & -9.1(2)  & -10.3\\
	    &  -8.9(2) & -9.4         &   -8.9     &  -9.0   & -7.5       &  -6.7 & -9.4   & -8.9  & -9.9 \\
	    &           &       &        &         &        &   &    &    & \\
FGK         &  -9.8    & -9.6(2) & -9.3  & -9.2    &  ...  & ...  &  -9.3  & -9.7  &  -9.8 \\
            &  -9.7    & -9.5(Var A)   &  -9.2  &  -9.1   &  ...   & ...   & -8.8  & -9.6 & -9.3 \\
	    &  -9.5    & -9.2(3) & -8.9(3) & -8.1     &  ...  & ...  & -8.5 & -9.4   & -9.2 \\
	   &           &       &        &        &          &  &  &   &  \\ 
RSGs       & -9.4(2)   & -9.6  & -9.6(MOH-G64)  &  -9.2  &  -9.2  &  -9.4 & -8.5 & -9.5:  &  -9.9: \\
        &  -9.3     & -9.5(2) & -9.1    &   -9.1  & -9.1   &  -8.8 & -8.4(2) 
 & -9.2(2) &  -9.5 \\   
 	   &  -9.1(2)  &  -9.4(2) & -9.0   &  -8.9   & -8.7    &  -8.7 & ...  & 
   ...  & -9.2 \\
\bottomrule
\end{tabular}
\end{table} 


Some comments with respect to the data in this Table are helpful. It is clear that data for some of the
spectral type groups is lacking or incomplete such as for the YSGs (FGK) in NGC 6822 and IC 1613. Those stars are undoubtedly present, but have just not been identified in the published surveys, and they may also be of somewhat lower
luminosity in those two galaxies. A survey to identify and classify the hotter supergiants in M101 has not been completed, and although a survey identifying RSGs in NGC 2403 exists \citep{Zickgraf} confirming spectroscopy and photometry is lacking. In general,  the numbers for the hot supergiant group (O9.5 - B5) may not include the most luminous members because many of these stars are in crowded regions.

The six Local Group galaxies in Table 1 present a diverse group of galaxy types with a wide range of
luminosities.  The well known dependence of the most luminous ``blue'' stars on the parent galaxy is especially notable  for the evolved, post-main sequence A-type supergiants in the two lowest luminosity
and lowest mass galaxies, NGC 6822 and IC 1613.  Together with the SMC, these are also the galaxies with the
lowest metallicity, but the heavy element abundance in NGC 6822 however, is 
intermediate between the SMC and LMC. Reduced metallicity was also expected to 
reduce the opacity in the stellar atmospheres and increase their 
absolute visual magnitudes but the data for these galaxies showed the opposite effect. 
So although metallicity  undoubtedly plays a role in stellar evolution, the luminosity and mass of the parent galaxy is the primary determinant, to the first order, for the luminosities of the most luminous stars. This is
a size of sample effect. Assuming a similar slope for the mass function, the most massive galaxies will
have a larger
progenitor population of massive stars, and consequently,  at any given time,  we will therefore be more
likely to
observe their most luminous, evolved counterparts.

We also see another effect with this group of galaxies 
  related to the galaxy type.  M31 (Sb I-II) and M81 (Sb I -II), both
 high luminosity spirals, have lower luminosity evolved A-type blue stars compared with the Sc-type spirals and Magellanic irregulars. This second effect very likely reflects a dependence on the lower star formation rate, not mass, in the Sb spirals. Both of these spirals have above solar metallicites, although  M81 
 has a metallicity gradient in its disk like  the Milky Way and the Sc spirals M33 and M101, and  solar-type abundances in the outer parts.

This degeneracy, wherein the luminosity of the most luminous blue stars are dependent on
the host galaxy, complicated their use as direct distance indicators. The most luminous  red supergiants however, exhibit a nearly constant upper luminosity, related to the Humphreys-Davidson  limit, over a wide range of galaxy types and luminosities. The presence of extensive circumstellar dust and uncertain reddening in the most luminous RSGs though limits their
usefulness as distance indicators in the visual, but their luminosities in the infrared need to be further
studied.  Kudritzki \citep{KudFGLR} has introduced a new distance determination method
for B and A-type supergiants, the Flux-weighted Gravity-Luminosity Relation (FGLR) that
depends on quantitative analysis  of the Balmer lines measured in low resolution spectra with good S/N ratio. It has  been successfully applied to  several nearby galaxies \citep{Urbaneja,U,Kud}, demonstrating that the visually  brightest stars, the A-type supergiants, have
potential as distance indicators at very large distances.

\section{The Complex Upper HR Diagram}

The upper luminosity boundary to the HR Diagram, or Humphreys-Davidson Limit, 
complicates our understanding of massive star evolution. Above some initial mass, $\approx$ 40 - 50 M$_{\odot}$, the stars do not evolve across the HR Diagram to become RSGs. 
Their evolution to cooler temperatures is most likely halted by high mass loss 
events and their  subsequent 
 proximity to the Eddington Limit \citep{HD16,Ulmer}. They then evolve back to warmer temperatures, now
  increasingly subject to instabilities, before their terminal state. Consequently, they have a different evolutionary path and different mass loss histories than their somewhat lower mass counterparts in the same galaxy or even the same cluster.  
  These lower mass supergiants, $\approx$ 9 or 10 up to 30 -- 40  M$_{\odot}$ or so, evolve across the HR Diagram , becoming red supergiants 
  which alters their interior structure and with 
  enhanced mass loss.  Some may also evolve back to a warmer state, perhaps as warm hypergiants, to become LBVs, or B[e] supergiants, where their 
  final fate is most likely as core-collapse supernovae.   

Consequently, the upper HR diagram is populated by a diversity of evolved, luminous and 
variable stars of different types that challenge our understanding of their
physics, evolution and eventual fate. Many of them are distinguished by their 
emission line spectra, and evidence for stellar winds and mass loss. 
In addition, some of them exhibit periods of enhanced mass loss, such as 
the LBVs/S Dor variables and the warm and cool hypergiants with 
their resolved ejecta. In addition, Wolf-Rayet stars of various types, 
Oe and Of stars, the B[e] supergiants, and the Fe II emission line stars occupy the same parts of the HR Diagram. They may or may not be related. 
They could be stars of similar initial mass but in different stages of their 
evolution or have experienced different mass loss histories, and some may be binaries. This diversity 
is one of the challenges to understanding massive stars, their evolution, and 
eventual fate.

The HR Diagrams below for M31 and M33 show the distribution in the 
luminosity-temperature plane of three of these mass losing classes of stars, the LBVs, B[e] supergiants and the warm hypergiants.  

\begin{figure}[H] 
\begin{subfigure}[b]{0.5\textwidth}
\includegraphics[width=\textwidth]{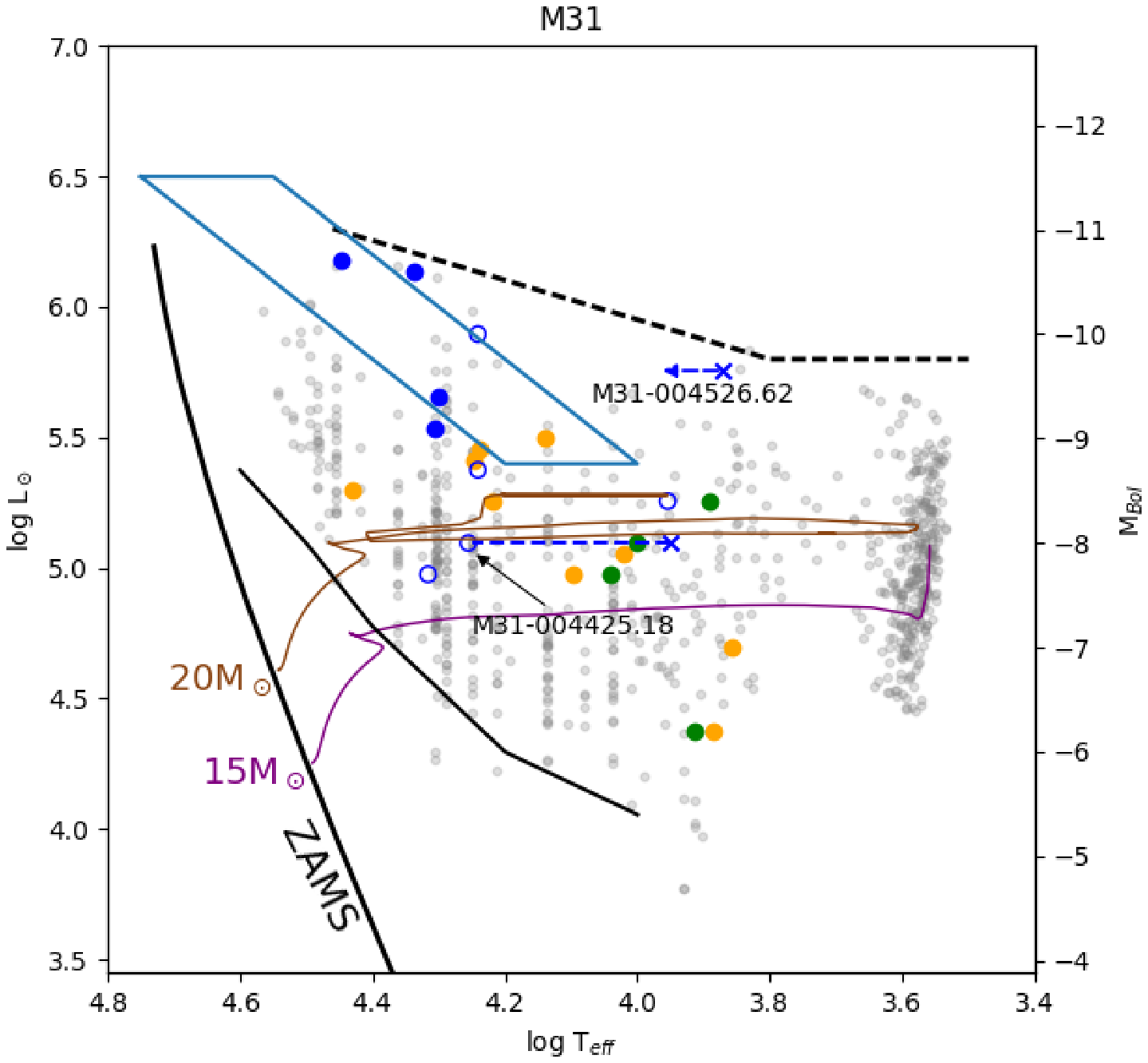}
\end{subfigure}
\begin{subfigure}[b]{0.5\textwidth}
\includegraphics[width=\textwidth]{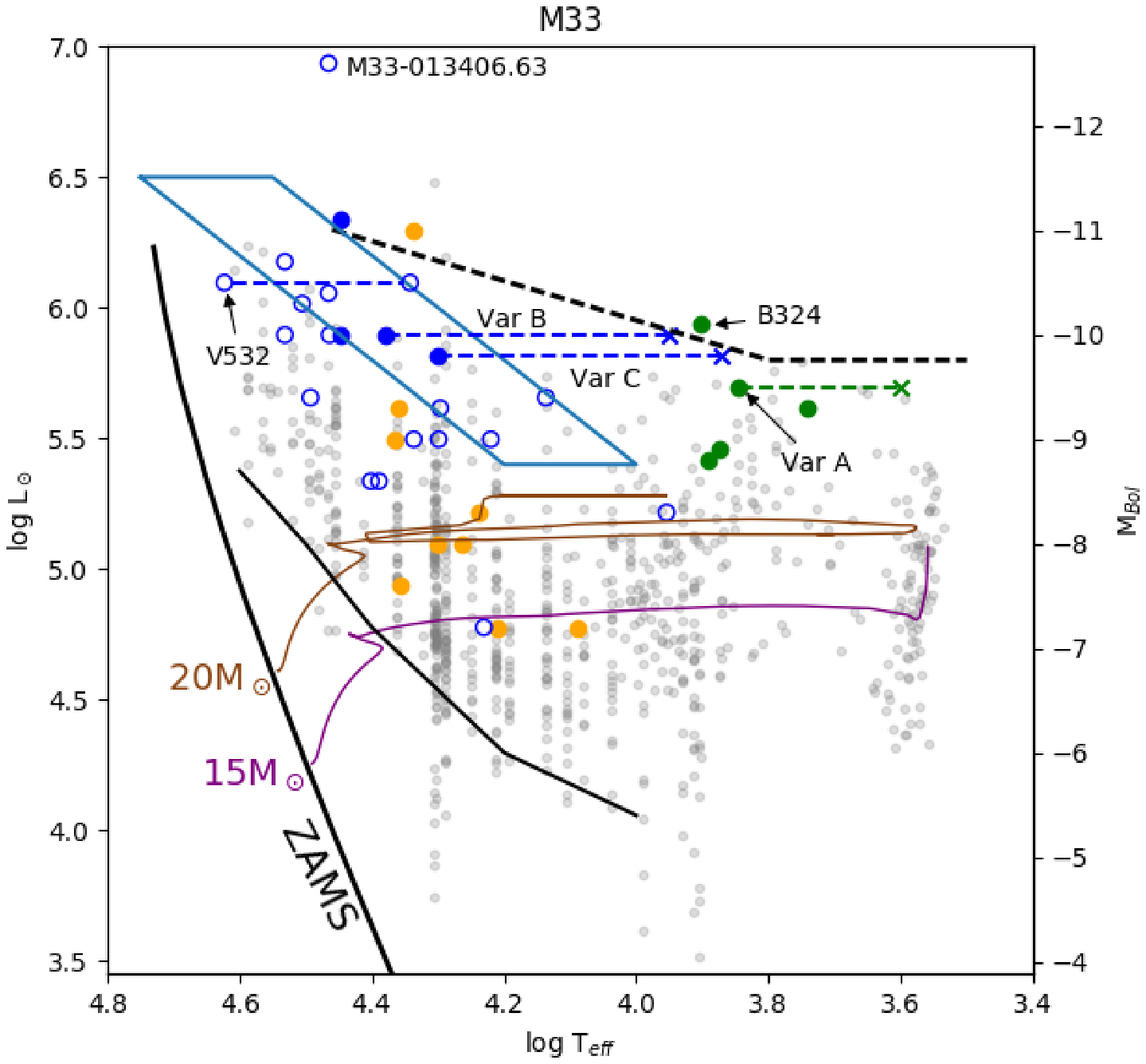}
\end{subfigure}
\caption{The schematic HR Diagrams  for M31 and M33 showing the positions of the confirmed LBVs
and candidate LBVs shown respectively, as filled and open blue circles,
the warm  hypergiants as green circles and the B[e] supergiants as orange
circles.  
The LBV transits during their high mass loss state are shown as dashed 
blue lines. The LBV/S Dor instability strip is outlined in blue. The 15 and 20 M$_{\odot}$ tracks are from  \citet{Ekstrom}
with rotation are shown to provide a reference for the lower mass B[e]sgs which are probable rotators. (Higher mass tracks are not shown due 
to crowding.)  The supergiant population  is shown in the background in light gray. Reproduced from \citet{RMH2017b} }
\end{figure}

One of the outstanding questions is the final fate of the most massive stars. It used to be simply assumed that all stars much above initial masses of 10 M$_{\odot}$ or so, would end their brief lives as some kind of supernova,  but their final stage as core-collapse SNe is now in question. Smartt \cite{Smartt2009,Smartt2015} has 
suggested an upper mass 
limit of $\approx$ 18 M${_\odot}$ for the red supergiant progenitors of 
the Type II-P SNe, while \citet{Jennings} find
a lack of massive supernova progenitors in M31 and M33 and suggest an upper mass 
of 35-45 M${_\odot}$. So, do the most massive stars collapse directly to 
black holes instead, long suspected for extreme, very massive stars like eta Carinae? 

In addition, the supernova surveys have identified numerous non-terminal giant
eruptions, in which the object greatly increases its total luminosity possibly 
expelling several solar masses and the star survives. Some of these events are 
confused with true 
SNe and thus have been  called ``supernova impostors''.  
This is a  diverse group of objects with a range of 
luminosities and possible progenitors. A few impostors appear to be normal 
LBV/S Dor variables in their eruptive or maximum light state. Most 
are giant eruptions, possibly similar to $\eta$ Car \citep{HD94} from evolved massive stars, while some  
are red transients (ILRTs, \citep{Bond}) from a lower mass population.   
The  origin of the instability in these giant eruptions is unknown, but proximity to the 
 Eddington Limit is  crucial \citep{Davidson}.  
It isn't known what role these high mass loss events may play in the final stages of  massive star evolution or their relation to 
the evolved massive star population and to other stars with instabilites like 
the  LBVs and the warm hypergiants.     

Thus we observe a complex upper HR diagram with different evolutionary paths dependent on 
initial mass, and several types of evolved  stars
not only experiencing continuous mass loss, but also  high mass loss events.
The study  of luminous stars in the nearer galaxies 
  provides us  with an improved census of these evolved stars, their
  relative numbers, physical properties 
  and behavior, and clues to their evolutionary state and possible  
  relationship to each other on the HR Diagram.  
 In this volume on luminous stars in nearby galaxies,  the reviews  
 focus on different examples of evolved massive stars; the luminous O and B-type stars in the Magellanic Clouds, LBVs, B[e] supergiants, the warm hypergiants, and the Wolf-Rayet stars.     

Although the chapters are about different types of evolved  massive stars, many will be in the same galaxies. We have therefore adopted the following distance moduli based on the Cepheid scale for these nearby galaxies for consistency 
and for ease of cross-referencing and comparison. 

\begin{table}[H]
\caption{Adopted Distance Moduli}
\centering
\begin{tabular}{ccc}
\toprule
\textbf{Galaxy}        & \textbf{Distance Modulus (mag)} &  \textbf{Comment}\\
\midrule
LMC         & 18.5                   & Cepheids  \\
SMC         & 18.9                  &  Cepheids      \\
M31         & 24.4                 & Cepheids \citep{M31Ceph}  \\
M33         & 24.5                 & Cepheids \citep{M33Ceph}   \\
NGC 6822    & 23.4                     &  Cepheids \citep{Feast2012} \\ 
IC 1613     & 24.3                 &    Cepheids \citep{IC1613Ceph}  \\ 
NGC 2403    & 27.5               & Cepheids \citep{Freedman}   \\ 
M81         & 27.8               & Cepheids \citep{Freedman} \\
M101      &  29.1               &  Cepheids \citep{Freedman}  \\  
\bottomrule
\end{tabular}
\end{table}


\reftitle{References}


\begin{thebibliography}{999}
\bibitem[Tammann \& Sandage(1968)]{TM}
Tammann, G. A. \&  Sandage, A. The Stellar Content and Distance of the Galaxy NGC 2403 in the M81 Group {\em ApJ} {\bf 1968}, {\em 151}, 825-860, doi:10.1086/149487.

\bibitem[Hubble \& Sandage(1952)]{HS}
Hubble, E. \& Sandage, A. The Brightest Variable Stars in Extragalactic Nebulae. I. M31 and M33. {\em ApJ} {\bf 1953}, {\em 118}, 353-361, doi:10.1086/145764.

\bibitem[Cannon(1936)]{Cannon}
Cannon, A. J. The Henry Draper Extension {\em HA} {\bf 1936}, {\em 100}.

\bibitem[Cannon \& Pickering(1918-1924)]{HA} 
Cannon, A. J. \&  Pickering, E. C.  The Henry Draper Catalog {\em HA} {\bf 1918-1924}, {\em 91-99}.

\bibitem[Feast, Thackeray \& Wesselink(1960)]{Feast}
Feast, M. W., Thackeray, A. D., \& Wesselink, A. J. The Brightest Stars in the 
Magellanic Clouds {\em MNRAS} {\bf 1960}, {\em 121}, 337-385,  doi:10.1093mnras/121.4.33 

\bibitem[Henize(1956)]{Henize}
Henize, K. Catalogues of H$\alpha$-emission Stars and Nebulae in the Magellanic 
Clouds {\em ApJS} {\bf 1956}, {\em 2}, 315-344, doi: 10.1086/190025

\bibitem[Sanduleak(1968)]{Sanduleak68}
Sanduleak, N.  A finding list of proven or probable Small Magellanic Clouds members. 
{\em AJ} {\bf 1968}, {\em 73}, 246-250, doi:10.1086/110625. 

\bibitem[Sanduleak(1970)]{Sanduleak70}
Sandulaeak, N.  A deep objective-prism survey for Large Magellanic Cloud members
. {\em Contributions of the Cerro Tololo Interamerican Observatory} {\bf 1970}, 
{\em 89}. 

\bibitem[Fehrenbach \& Duflot(1970)]{Fehrenbach}
ehrenbach, C. \& Duflot, M. Large Magellanic Cloud. List of LMC members and lis
t of galactic stars. Charts for recognizing these stars on the maps published by
 the Smithsonian Institution Astrophysical Observatory. {\em A\&A Special Suppl.} 
 {\bf 1970} {\em 1}. 

\bibitem[Westerlund(1960)]{West60}
Westerlund, B. An infrared survey of the Magellanic Clouds. I. Four regions in the 
Large Cloud. {\em Uppsala Ann.} {\bf 1960}, Vol 4, No.7.

\bibitem[Westerlund(1961)]{West61}
Westerlund, B. Population I in the Large Magellanic Cloud. {\em Uppsala Ann.} {\bf 1961}, Vol 5, No 1, 1-28. 

\bibitem[Blanco, McCarthy \& Blanco(1980)]{Blanco80}
Blanco, V. M., McCarthy, M. F., \& Blanco, B. Carbon and late M-type stars in the 
Magellanic Clouds. {\em ApJ} {\bf 1980}, {\em 242}, 938-964, doi:10.1086/158527.

\bibitem[Blanco \& McCarthy(1983)]{Blanco83}
Blanco, V. M. \& McCarthy, M. F. The distribution of carbon and M-type giants in
 the Magellanic Clouds. {\em AJ} {\bf 1983}, {\em 88}, 1442-1457, doi:10.1086/113433. 

\bibitem[Sandage \& Tammann(1974)]{ST74}
Sandage, A. \& Tammann, G. A. Steps toward the Hubble constant. II. The brightest stars 
in late-type spiral galaxies. {\em ApJ} {\bf 1974}, {\em 191}, 603-621, doi:10.1086/153001.

\bibitem[Humphreys(1978)]{RMH78}
Humphreys, R. M. Studies of luminous stars in nearby galaxies. I. Supergiants and O stars in the Milky Way. {\em ApJS} {\bf 1978}, {\em 38}, 309-350, doi:10.1086/190559.

\bibitem[Humphreys(1973)]{RMH73}
Humphreys, R. M. Spectroscopic and photometric observations of luminous stars in
 Carina-Centaurus (l=282d - 305d. {\em A\&A} {\bf 1973}, {\em 9}, 85-96.

\bibitem[Humphreys(1975)]{RMH75}
Humphreys, R. M. Spectroscopic and photometric observations of luminous stars in
 the Centaurus-Norma (l = 305 - 340 ) section of the Milky Way.
  {\em A\&A Suppl.} {\bf 1975}, {\em 19}, 243 - 247.

\bibitem[Walborn(1972)]{Walborn72}
Walborn, N. R. Spectral classification of OB stars in both hemispheres and the absolute-magnitude calibration. {\em AJ} {\bf 1972}, {\em 77}, 312 - 318, doi:10.1086/111285. 

\bibitem[Humphreys, Strecker \& Ney(1972)]{RMH72}
Humphreys, R. M., Strecker, D. W., \& Ney, E. P. Spectroscopic and Photometric Observations of M Supergiants in Carina. {\em ApJ} {\bf 1972}, {\em 172}, 75-88, doi:10.1086/151329

\bibitem[Humphreys \& Ney(1974)]{RMH74}
Humphreys, R. M. \& Ney, E. P. Visual and infrared observations of late-type supergiants in the southern sky. {\em ApJ} {\bf 1974}, {\em 194}, 623-628, doi: 10.1086/153282.

\bibitem[Humphreys(1979a)]{RMH79}
Humphreys, R. M. Studies of luminous stars in nearby galaxies. II - M supergiants in the Large Magellanic Cloud. {\em ApJS} {\bf 1979}, {\em 39}, 389-403, doi:10.1086/190578. 


\bibitem[Ardeberg et al.(1972)]{Ardeberg}
Ardeberg, A., et al. Spectrographic and photometric observations of supergiants and foreground stars in the direction of the Large Magellanic Cloud. {\em A\&A Suppl.} {\bf 1972}, {\em 6}, 249-309.

\bibitem[Brunet et al.(1975)]{Brunet}
Brunet, J-P., et al. Studies of the LMC stellar content. I. A catalogue of 272 new O-B2 stars. {\em A\&A Suppl.} {\bf 1975}, {\em 21}, 109 - 136. 


\bibitem[Walborn(1977)]{Walborn77}
Walborn, N. R. Spectral classification of O and B0 supergiants in the Magellanic Clouds. {\em ApJ} {\bf 1977}, {\em 215},  53-57, doi:10.1086/155334

\bibitem[Humphreys \& Davidson(1979)]{HD79}
Humphreys, R. M. \& Davidson, K. Studies of luminous stars in nearby galaxies. 
III - Comments on the evolution of the most massive stars in the Milky Way and the 
Large Magellanic Cloud. {\em ApJ} {\bf 1979}, {\em 232}, 409-420, doi:10.1086/157301.

\bibitem[Humphreys(1979b)]{RMH79b}
Humphreys, R. M. Studies of luminous stars in nearby galaxies. IV - Baade's field 
IV in M31. {\em ApJ} {\bf 1979}, {\em 234}, 854-860, doi:10.1086/157566.  

\bibitem[Humphreys(1980a)]{RMH80a}
Humphreys, R. M. Studies of luminous stars in nearby galaxies. V - The local group 
irregulars NGC 6822 and IC 1613. {\em ApJ} {\bf 1980}, {\em 238}, 65-78, 10.1
086/157958.  


\bibitem[Humphreys \& Sandage(1980)]{HS80}
Humphreys, R. M. \& Sandage, A. On the stellar content and structure of the spiral 
Galaxy M33. {\em ApJS}, {\bf 1980}, {\em 44}, 319-381, doi:10.1086/190696.



\bibitem[Humphreys(1980b)]{RMH80b}
Humphreys, R. M. Studies of luminous stars in nearby galaxies. VI - The brightest 
supergiants and the distance to M33. {\em ApJ} {\bf 1980}, {\em 241},  587-597
, doi: 10.1086/158373.  

\bibitem[Humphreys(1983)]{RMH83}
Humphreys, R. M. Studies of luminous stars in nearby galaxies. VIII - The Small 
Magellanic Cloud. {\em ApJ} {\bf 1983}, {\em 265}, 176-193, doi: 10.1086/160662.

\bibitem[Humphreys(1975)]{RMH75_HS}
Humphreys, R. M. The spectra of AE Andromedae and the Hubble-Sandage variables in 
M31 and M33. {\em ApJ} {\bf 1975}, {\em 200}, 426-429, doi: 10.1086/153806 

\bibitem[Humphreys(1978)]{RMH78_HS}
Humphreys, R. M. Luminous variable stars in M31 and M33. {\em ApJ} {\bf 1978}, {\em 219}, 445-451, doi: 10.1086/155797

\bibitem[Chiosi, Nasi, \& Sreenivasan(1977)]{Chiosi}
Chiosi, C., Nasi, E., \& Sreenivasan, S. R. Massive stars evolution with mass-loss. 
{\em A\&A} {\bf 1978}, {\em 63}, 103-124. 

\bibitem[de Loore, De Greve, \& Lamers(1977)]{deLoore} 
de Loore, C., De Greve, J. P., \& Lamers, H. J. G. L. M. Evolution of massive stars with mass loss by stellar wind.  {\em A\&A} {\bf 1977}, {\em 61}, 251-259.

\bibitem[Maeder(1980)]{Maeder80}
Maeder, A. The most massive stars in the Galaxy and the LMC - Quasi-homogeneous 
evolution, time-averaged mass loss rates and mass limits {\em A\&A} {\bf 1980}, 
{\em 92}, 101-110 

\bibitem[Maeder(1981)]{Maeder81}
Maeder, A. Grids of evolutionary models for the upper part of the HR diagram. Mass 
loss and the turning of some red supergiants into WR stars. {\em A\&A} {\bf 1
981}, {\em 102}, 401-410.

\bibitem[Elias, Frogel, \& Humphreys(1985)]{Elias}
Elias, J. H., Frogel, J. A., \& Humphreys, R. M. M supergiants in the Milky Way 
and the Magellanic Clouds :colors, spectral types, and luminosities. {\em ApJS} 
{\bf 1985}, {\em 57}, 57-131, doi:10.1086/190997.

\bibitem[Skrutskie et al.(2006)]{Skrutskie}
Skrutskie, M. F. et al. The Two Micron All Sky Survey (2MASS). {\em AJ} {\bf 2006}, {\em 131}, 1163-1183, doi: 10.1086/498708 

\bibitem[Bonanos et al.(2009)]{Bonanos09} 
Bonanos, A. et al. Spitzer SAGE Infrared Photometry of Massive Stars in the Large 
Magellanic Cloud  {\em AJ} {\bf 2009}, {\em 138}, 1003-1021, doi: 10.1088/0004
-6256/138/4/1003. 

\bibitem[Bonanos et al.(2010)]{Bonanos10}
Bonanos, et al. Spitzer SAGE-SMC Infrared Photometry of Massive Stars in the Small 
Magellanic Cloud. {\em AJ} {\bf 2010}, {\em 140}, 416-429, doi: 10.1088/0004-
6256/140/2/416. 

\bibitem[Mould et al.(2008)]{Mould}
Mould, J. et al. A Point-Source Survey of M31 with the Spitzer Space Telescope. 
{\em ApJ} {\bf 2008}, {\em 687}, 230-241, doi: 10.1086/591844

\bibitem[McQuinn et al.(2007)]{McQuinn}
McQuinn, K. B. W., et al. The M33 Variable Star Population Revealed by Spitzer. 
{\em ApJ} {\bf 2007}, {\em 664}, 850-861, doi: 10.1086/519068. 

\bibitem[Massey et al.(2006)]{Massey}
Massey, P., et al.  A Survey of Local Group Galaxies Currently Forming Stars. I.
 UBVRI Photometry of Stars in M31 and M33. {\em AJ} {\bf 2006}, {\em 131}, 2478-
  2496, doi: 10.1086/503256.


\bibitem[Dalcanton, et al.(2012)]{Dalcanton} 
Dalcanton, J. J. et al. The Panchromatic Hubble Andromeda Treasury. {\em ApJS} {\bf 2012}, {\em 200}, 1-37, doi: 10.1088/0067-0049/200/2/18. 

\bibitem[Grammer, Humphreys, \& Gerke(2015)]{Grammer}
Grammer, S. H., Humphreys, R. M., \& Gerke, J. The Massive Star Population in M101. III. Spectra and Photometry of the Luminous and Variable Stars. {\em 149} {\bf 2015}, 
{\em 149:152}, 1-15, doi: 10.1088/0004-6256/149/5/152. 

\bibitem[Humphreys et al.(2019)]{RMH2019}
Humphreys, R. M., Stangl, S., Gordon, M. S., et al. Luminous and Variable Stars 
in NGC 2403 and M81. {\em AJ} {\bf 2019}, {\em 157:22}. 1-16, doi: 10.3847/1538-3881/aaf1ac. 

\bibitem[Humphreys(1979c)]{RMH_RSG}
Humphreys, R. M. M supergiants and the low metal abundances in the Small Magellanic Cloud. {\em ApJ} {\bf 1979}, {\em 231}, 384-389, doi: 10.1086/157201. 

\bibitem[Massey \& Olsen(2003)]{MasseyOL}
Massey, P. \& Olsen, K. A. G. The Evolution of Massive Stars. I. Red Supergiants in 
the Magellanic Clouds. {\em AJ} {\bf 2003}, {\em 126}, 2867-2886, doi: 10.1086/379558.  


\bibitem[Vink, de Koter \& Lamers(2001)]{Vink}
Vink, J. S., de Koter, A., \& Lamers, H. J. G. L. M. Mass-loss predictions for O
 and B stars as a function of metallicity. {\em A\&A} {\bf 2001}, {\em 369}, 574
 -588, doi: 10.1051/0004-6361:20010127.

\bibitem[Mokiem et al.(2007)]{Mokiem}
Mokiem, M. R. et al. The empirical metallicity dependence of the mass-loss rate 
of O- and early B-type stars. {\em A\&A} {\bf 2007}, {\em 473}, 603-614, doi: 10.1051/0004-6361:20077545.   

\bibitem[Crowther et al.(2002)]{Crowther}
Crowther, P. A., Hillier, D. J., Evans, C. J. et al. Revised Stellar Temperatures for Magellanic Cloud O Supergiants from Far Ultraviolet Spectroscopic Explorer and Very Large Telescope UV-Visual Echelle Spectrograph Spectroscopy. {\em ApJ} {\bf 2002}, {\em 579}, 774-799, doi: 10.1086/342877 

\bibitem[Hillier et al.(2003)]{Hillier}
Hillier, D. J., Lanz, T., Heap, S. R., et al. A Tale of Two Stars: The Extreme O7 Iaf+ Supergiant AV 83 and the OC7.5 III((f)) star AV 69. {\em ApJ} {\bf 2003}, {\em 588}, 1039-1063, doi: 10.1086/374329

\bibitem[Fullerton et al.(2006)]{Fullerton}
Fullerton, A. W., Massa, D. L. \& Prinja, R. K. The Discordance of Mass-Loss Estimates for Galactic O-Type Stars. {\em ApJ} {\bf 2006}, {\em 637}, 1025-1039, doi: 10.1086/498560 


\bibitem[Neugent et al.(2010]{Neugent2010}
Neugent, K. F., Massey, P., Skiff, B., et al. Yellow Supergiants in the Small 
Magellanic Cloud: Putting Current Evolutionary Theory to the Test. {\em ApJ} 
{\bf 2010}, {\em 719},  1784-1795, doi: 10.1088/0004-637X/719/2/1784 

\bibitem[Neugent et al.(2012)]{Neugent2012}
Neugent, K. F., Massey, P., Skiff, B., et al. Yellow and Red Supergiants in the 
Large Magellanic Cloud. {\em ApJ} {\bf 2012}, {\em 749}, 177, 12pp, doi: 10.1088
/0004-637X/749/2/177 

\bibitem[Humphreys et al.(2013)]{RMH2013}
Humphreys, R. M., Davidson, K., Grammer, S., et al. Luminous and Variable Stars 
in M31 and M33. I. The Warm Hypergiants and Post-red Supergiant Evolution. {\em 
ApJ} {\bf 2013}, {\em 773}, 46, 11pp, doi: 10.1088/0004-637X/773/1/46


\bibitem[Humphreys et al(2014)]{RMH2014}
Humphreys, R. M., Weis, K., Davidson, K., et al. Luminous and Variable Stars in 
M31 and M33. II. Luminous Blue Variables, Candidate LBVs, Fe II Emission Line St
ars, and Other Supergiants. {\em ApJ}, {\bf 2014}, {\em 790}, 48, 21pp, doi: 10.
1088/0004-637X/790/1/48

\bibitem[Massey et al.(2016b)]{Massey2016b}
Massey, P., Neugent, K. F., \& Smart, B. M. A Spectroscopic Survey of Massive Stars 
in M31 and M33. {\em AJ}, {\bf 2016}, {\em 152}, 62, 16pp, doi: 10.3847/0004
-6256/152/3/62

\bibitem[Humphreys et al.(2017a)]{RMH2017a}
Humphreys, R. M., Gordon, M. S., Martin, J. C.; et al. Luminous and Variable Stars 
in M31 and M33. IV. Luminous Blue Variables, Candidate LBVs, B[e] Supergiants
, and the Warm Hypergiants: How to Tell Them Apart. {\em ApJ}, {\bf 2017}, {\em 836}, 
64, 13pp, doi: 10.3847/1538-4357/aa582e

\bibitem[Humphreys et al.(2017b)]{RMH2017b}
Humphreys, R. M.; Davidson, K., Hahn, D., et al. Luminous and Variable Stars in
M31 and M33. V. The Upper HR Diagram. {\em ApJ} {\bf 2017}, {\em 844}, 40, 9pp,
doi: 10.3847/1538-4357/aa7cef

\bibitem[Massey et al.(2009)]{Massey2009} 
Massey, P., Silva, D. R., Levesque, E. M., et al.  Red Supergiants in the Andromeda 
Galaxy (M31). {\em AJ} {\bf 2009}, {\em 703}, 420-440, doi: 10.1088/0004-637
X/703/1/420 

\bibitem[Drout et al.(2009)]{Drout2009}
Drout, M. R., Massey, P., Meynet, G., et al.  Yellow Supergiants in the Andromeda 
Galaxy (M31) {\em ApJ}, {\bf 2009}, {\em 703}, 441-460, doi: 10.1088/0004-637X
/703/1/441

\bibitem[Drout et al.(2012)]{Drout2012}
Drout, M. R., Massey, P., Meynet, G. et al. The Yellow and Red Supergiants of M33. {\em ApJ} {\bf 2012}, {\em 750}, 97, 22pp, doi: 10.1088/0004-637X/750/2/97 


\bibitem[Gordon et al.(2016)]{Gordon}
Gordon, M. S., Humphreys, R. M., \& Jones, T. J. Luminous and Variable Stars in 
M31 and M33. III. The Yellow and Red Supergiants and Post-red Supergiant Evolution. 
{\em ApJ}, {\bf 2016}, {\em 825}, 50, 17pp, doi: 10.3847/0004-637X/825/1/50

\bibitem[Massey \& Evans(2016a)]{Massey2016a}
Massey, P. \& Evans, K. A. The Red Supergiant Content of M31. {\em ApJ} {\bf 2016}, 
{\em 826}, 224, 9pp, doi: 10.3847/0004-637X/826/2/224

\bibitem[Massey et al.(1995)]{Massey95}
Massey, P., et al. Hot, Luminous Stars in Selected Regions of NGC 6822, M31, and
 M33. {\em AJ} {\bf 1995}, {\em 110}, 2715-2745, doi: 10.1086/117725

\bibitem[Massey,(1998)]{Massey98}
Massey, P. Evolved Massive Stars in the Local Group. I. Identification of Red 
Supergiants in NGC 6822, M31, and M33. {\em ApJ} {\bf 1998}, {\em 501}, 153-174, doi: 10.1086/305818

\bibitem[Bresolin et al.(2007)]{Bresolin07}
Bresolin, F., et al. VLT Spectroscopy of Blue Supergiants in IC 1613. {\em ApJ} 
{\bf 2007}, {\em 671}, 2028-2039, doi: 10.1086/522571


\bibitem[Humphreys et al.(1986)]{RMH86}
Humphreys, R M., Aaronson, M., Liebofsky, M,  et al. The luminosities of M supergiants 
and the distances to M 101, NGC 2403, and M 81. {\em AJ}, {\bf 1986}, {\em 91}, 808-821, doi: 10.1086/114061

\bibitem[Humphreys \& Aaronson(1987)]{RMH87}
Humphreys, R. M. \& Aaronson, M. The Visually Brightest Early-Type Supergiants in 
the Spiral Galaxies NGC 2403, M81, and M101  {\em AJ} {\bf 1987}, {\em 94}, 1156-1169, doi: 10.1086/114553 

\bibitem[Zickgraf \& Humphreys(1991]{Zickgraf}
Zickgraf, F-J \& Humphreys, R. M A Stellar Content Survey of NGC 2403 and M81. {\em AJ} {\bf 1991}, {\em 102}, 113-133 doi: 10.1086/115860


\bibitem[Kudritzki et al.(2012)]{Kud}
Kudritzki, R-P, et al. Quantitative Spectroscopy of Blue Supergiant Stars in the
 Disk of M81: Metallicity, Metallicity Gradient, and Distance. {\em ApJ} {\bf 2012}, {\em 747}, 15, 19 pp., doi: 10.1088/0004-637X/747/1/15

\bibitem[Kudritzki et al.(2008)]{KudFGLR}
Kudritzki, R-P, et al.  Quantitative Spectroscopy of 24 A Supergiants in the Sculptor Galaxy NGC 300: Flux-weighted Gravity-Luminosity Relationship, Metallicity, and Metallicity Gradient. {\em apJ} {\bf 2008}, {\em 681}, 269-289, doi: 10.1086/588647

\bibitem[Urbaneja et al.(2008)]{Urbaneja}
Urbaneja, M. A. The Araucaria Project: The Local Group Galaxy WLM—Distance and Metallicity from Quantitative Spectroscopy of Blue Supergiants. {\em ApJ} {\bf 2008}, {\em 684}, 118-135, doi: 10.1086/590334 

\bibitem[U. et al.(2009)]{U}
U., V. et al. A New Distance to M33 Using Blue Supergiants and the FGLR Method. {\em ApJ} {\bf 2009}, {\em 704}, 1120-1134, doi: 10.1088/0004-637X/704/2/1120

\bibitem[Humphreys et al(2016)]{HD16}
Humphreys, R. M., Weis, K., Davidson, K., \& Gordon, M. S. On the Social Traits of Luminous Blue Variables. {\em ApJ} {\bf 2016}, {\em 825},  64, 15 pp., doi: 10.3847/0004-637X/825/1/64 

\bibitem[Ulmer \& Fitzpatrick(1998)]{Ulmer}
Ulmer, A. \& Fitzpatrick, E. L. Revisiting the Modified Eddington Limit for Massive Stars. {\em ApJ}  {\bf 1998}, {\em 504}, 200-206, doi: 10.1086/306048 

\bibitem[Ekstrom, et al(2012)]{Ekstrom}
Ekstrom, S., Georgy, C., Eggenberger, P., et al. Grids of stellar models with rotation. I. Models from 0.8 to 120 M$_{\odot}$  at solar metallicity (Z = 0.014).
 {\em A\&A} {\bf 2012}, {\em 537}, A146, 18pp, doi: 10.1051/0004-6361/201117751

\bibitem[Smartt et al.(2009)]{Smartt2009}
Smartt, S. J., Eldridge, J. J., Crockett, R. M., et al.  The death of massive 
stars - I. Observational constraints on the progenitors of Type II-P supernovae.  
{\em MNRAS} {\bf 2009}, {\em 395}, 1409-1437, doi: 10.1111/j.1365-2966.2009.14506

 \bibitem[Smartt(2015)]{Smartt2015} 
Smartt, S. J. Observational Constraints on the Progenitors of Core-Collapse Supernovae: 
The Case for Missing High-Mass Stars. {\em PASA} {\bf 2015}, {\em 32} 22 pp, doi: 10.1017/pasa.2015.17

\bibitem[Jennings et al.(2014)]{Jennings}
Jennings, Z. G., et al. The Supernova Progenitor Mass Distributions of M31 and M33: Further Evidence for an Upper Mass Limit. {\em ApJ} {\bf 2014}, {\em 795}, 1
70, 11pp, doi: 10.1088/0004-637X/795/2/170

\bibitem[Humphreys \& Davidson(1994)]{HD94}
Humphreys, R. M. \& Davidson, K.  The Luminous Blue Variables: Astrophysical Geysers.{\em PASP} {\bf 1994}, {\em 106}, 1025-1051, doi: 10.1086/133478

\bibitem[Bond(2011)]{Bond}
Bond, H. E. Hubble Space Telescope Imaging of the Outburst Site of M31 RV. II. No Blue Remnant in Quiescence. {\em ApJ} {\bf 2011}, {\em 737}, 17, 6pp, doi: 10.1088/0004-637X/737/1/17

\bibitem[Davidson(2016)]{Davidson}
Davidson, K. Giant eruptions of very massive stars. In {\em Late Stages of Stell 
Evolution} {\bf 2016}, Journal of Physics: Conference Series {\em 728} doi: 10.1088/1742-6596/728/2/022008 

\bibitem[Riess et al.(2012)]{M31Ceph}
Riess, A. G., Fliri, J., \& Valls-Gabaud, D. Cepheid Period-Luminosity Relations
 in the Near-infrared and the Distance to M31 from the Hubble Space Telescope Wide Field Camera 3. {\em ApJ} {\bf 2012}, {\em 745},  156, 6pp, doi: 10.1088/0004-637X/745/2/156 

\bibitem[Scowcroft et al.(2009)]{M33Ceph}
Scowcroft, V., Bersier, D., Mould, J. R ., \& Wood, P. R.  The effect of metallicity 
on Cepheid magnitudes and the distance to M33. {\em MNRAS} {\bf 2009}, {\em 396} 1287-1296, doi: 10.1111/j.1365-2966.2009.14822.x

\bibitem[Feast et al.(2012)]{Feast2012}
Feast, M. W. et al. The Cepheid distance to the Local Group galaxy NGC 6822. {\em MNRAS}, {\bf 2012}, {\em 421},  2998-3003, doi: 10.1111/j.1365-2966.2012.20525.x

\bibitem[Scowcroft et al.(2013)]{IC1613Ceph}
Scowcroft, V., et al. The Carnegie Hubble Program: The Infrared Leavitt Law in IC 1613. 
{\em ApJ} {\bf 2013}, {\em 773}, 106, 13 pp., doi: 10.1088/0004-637X/773/2/106 

\bibitem[Freedman et al.(2001)]{Freedman}
Freedman, W. L., Madore, B. F., Gibson, B. K. et al. Final Results from the Hubble Space Telescope Key Project to Measure the Hubble Constant. {\em ApJ} {\bf 20
01},  {\em 553},  47-72,  doi: 10.1086/320638

\end{thebibliography}
\end{document}